\documentstyle[11pt,aaspp4]{article}

\newcommand{\NH}{{$N_{\rm H}$}}
\newcommand{\LX}{{$L_{\rm X}$}}
\newcommand{\LB}{{$L_{\rm B}$}}
\newcommand{\LFIR}{{$L_{\rm FIR}$}}
\newcommand{\LHa}{$L_{{\rm H} \alpha}$}
\newcommand{\LOIII}{$L_{\rm [OIII]}$}
\newcommand{\FX}{{$F_{\rm X}$}}

\newcommand{\eps}{ergs s$^{-1}$}
\newcommand{\pcm}{cm$^{-2}$}

\newcommand{\asca}{{\it ASCA}}
\newcommand{\Einstein}{{\it Einstein}}
\newcommand{\rosat}{{\it ROSAT}}
\newcommand{\HST}{{\it HST}}

\newcommand{\gtsima}{$\; \buildrel > \over \sim \;$}
\newcommand{\simgt}{\lower.5ex\hbox{\gtsima}}
\newcommand{\ltsima}{$\; \buildrel < \over \sim \;$}
\newcommand{\simlt}{\lower.5ex\hbox{\ltsima}}

\slugcomment{To appear in {\it The Astrophysical Journal.}}

\lefthead{Terashima et al.}
\righthead{Iron K Emission from NGC 4579}

\begin{document}

\title{Detection of an iron K Emission Line\\ from the LINER NGC 4579}

\author{Yuichi Terashima, Hideyo Kunieda, and Kazutami Misaki}
\affil{Department of Physics, Nagoya University, Chikusa-ku, Nagoya 464-8602,
Japan}

\author{Richard F. Mushotzky, Andrew F. Ptak\altaffilmark{1}, and
Gail A. Reichert}
\affil{Laboratory for High Energy Astrophysics, Code 662, NASA/Goddard Space
Flight
Center, Greenbelt, MD 20771}


\altaffiltext{1}{present address: Department of Physics, 
Carnegie Mellon University, 
5000 Forbes Ave., Pittsburgh, PA 15213}

\begin{abstract}

We present the results of
an {\asca} observation of the LINER NGC~4579. A point-like X-ray
source is detected at the nucleus with a 2--10~keV luminosity of
$1.5\times10^{41}$~{\eps}assuming a distance of 16.8~Mpc. The X-ray
spectrum is represented by a combination of a power-law with a
photon index of $\sim$1.7 and soft thermal component with
$kT\sim$0.9~keV. An iron K emission line is detected at
$6.73\pm0.13$~keV (rest frame) with an equivalent width of
$490^{+180}_{-190}$~eV and is statistically significant at more than
99.9\% confidence. The line center energy is consistent with
Helium-like iron and is significantly higher than 6.4~keV which is
expected from fluorescence by "cold" (or a lower ionization state of)
iron. The iron line profile shows no significant red tail in contrast
to Seyfert 1 galaxies although the statistics are limited. The line center
energy, equivalent width, and profile are consistent with an origin in
an ionized accretion disk. However the large mass accretion rate
necessary to ionize the accretion disk is not consistent with the
observed luminosity and normal accretion models.

\end{abstract}


\section{Introduction}

 Recent optical spectroscopic surveys have shown that there are many
active galactic nuclei (AGNs) in nearby galaxies and about 40\% of
bright galaxies are classified as Seyfert galaxies or LINERs (Low
Ionization Nuclear Emission-line Regions; \cite{hec80}) (Ho,
Filippenko, \& Sargent 1997a). The luminosity of these objects are
rather low compared to previously known AGN with a median value of the
H$\alpha$ luminosity being only $2\times10^{39}$ {\eps} in the sample
of Ho et al (1997a). Such objects (low luminosity AGNs;
hereafter LLAGNs) are important for investigating the physics of AGN
under an extreme condition, i.e. very low luminosity. X-ray
observations probe the innermost regions of AGNs and specifically the
iron K line provides information on the ionization state, density,
and motion of matter very close to the central energy source.

{\asca} observations of Seyfert 1 galaxies revealed that as a class
these objects have a broad iron K line with a profile skewed to lower
energies, thought to be caused by the reprocessing of the
continuum by a relativistic accretion disk (e.g., \cite{tan95}, Nandra
et al. 1997a). The center energy of the iron line from Seyfert 1
galaxies is consistent with 6.4~keV, which is expected from
fluorescence by neutral or lower ionization states of
($<$\ion{Fe}{16}) iron in a disk with an inclination of $<$ 30
degrees. In the Seyfert 1.9 galaxy IRAS ~18325--5926, a higher peak
energy of iron emission is seen, which is compatible with a highly-inclined
disk ($i=40-50^{\circ}$) origin (\cite{iwa96}). Highly-ionized iron emission
lines are detected from several radio-quiet quasars e.g.
E1821+643 (\cite{kii91}, \cite{yam97}) and PG~1116+215
(\cite{nan96}). Nandra et al. (1997b) studied the luminosity
dependence of the iron line profile in a large sample of AGN and found
that the center energy increases and the red-tail becomes weaker with
increasing luminosity. They attributed such behavior to an increasing
ionization of the accretion disk with increasing luminosity. Thus
X-ray measurements of iron emission lines are powerful diagnostic
tools of matter in the vicinity of the nucleus.

There are only a few observations of iron emission lines from low
luminosity AGNs ( {\LX} (2--10~keV) $\sim
10^{40}-10^{41}$~{\eps}). M81 (NGC~3031) with an X-ray luminosity of
{\LX} (2--10~keV) $\sim 2\times10^{40}$~{\eps} shows a broad iron line
centered at $\sim6.7$~keV with an equivalent width of $\sim 200$~eV.
This line center energy is significantly higher than Seyfert 1
galaxies and similar to luminous quasars. An iron line at 6.4~keV with
an equivalent width of $\sim300$~eV is detected from the low
luminosity Seyfert 1 galaxy NGC~5033 ({\LX} (2--10~keV) =
$2\times10^{41}$~{\eps}, \cite{te98b}) but only an upper limit on the
equivalent width of 
{\simlt} 300~eV is obtained for NGC~1097 ({\LX} (2--10~keV) =
$1\times10^{41}$~{\eps}, Iyomoto et al. 1996). Although strong iron
emission lines are also detected from M51 (= NGC~5194, Terashima et
al. 1998a), NGC~1365 and NGC~1386 (Iyomoto et al. 1997), the iron
lines in these objects are interpreted as being caused by reprocessed
emission from an obscuring tori and/or extended ionized scatterer
outside of our line of sight, i.e., these nuclei are heavily obscured.
Thus, at present, the number of LLAGNs with small intrinsic absorption
from which iron lines are detected is rather limited.

NGC~4579 (M58) is a Sab galaxy in the Virgo cluster of galaxies and
classified as a LINER or Seyfert 1.9 galaxy based on the optical
emission lines (\cite{ho97a}, \cite{kee83}, \cite{sta82}) and the
broad H$\alpha$ component, detected with a FWHM $\sim 2300$~km s$^{-1}$
(\cite{ho97b}). There exists a
flat-spectrum radio core (\cite{hum87}). An {\Einstein} HRI
observation showed the presence of an unresolved X-ray nucleus and the
X-ray flux was measured to be {\FX} = $7.9\times10^{-12}$~{\eps}
cm$^{-1}$ in the 0.2--4.0~keV band with the {\Einstein} IPC (\cite{fab92},
Halpern \& Steiner 1983) which corresponds to
the X-ray luminosity of $2.7\times10^{41}$~{\eps} (we assume a
distance of 16.8~Mpc (\cite{tul88}) throughout this paper). These facts
indicate the presence of a LLAGN in this galaxy. A recent
ultraviolet imaging observation by {\it Hubble Space Telescope}
({\HST}) Faint Object Camera (FOC) detected a point source at the
nucleus (\cite{mao95}). Its UV spectra were taken by {\HST} Faint
Object Spectrograph (FOS) and a featureless UV continuum is detected
as well as various emission lines. Comparison of the FOC and FOS data
also indicate a factor of 3.3 decrease of UV flux in 19 months. The
narrow UV emission lines are incompatible with shock excitation model
and a photoionization model is preferred (Barth et al. 1996). Several
broad UV emission lines are also detected. These UV results provide
further support for the presence of a LLAGN in NGC~4579.
On the other hand, \cite{mao98} estimated the ionizing photon
number by extrapolating the UV luminosity at 1300 A towards higher
energies and argued that the observed UV continuum is not sufficient to
explain the H$\alpha$ luminosity. They also suggest that emission from
AGNs is most prominent at energies higher than the UV. Measurements of an
X-ray flux and continuum slope provide information on the ionization
source in this LINER.  In this paper we report the detection of an Iron K
emission line centered at 6.7~keV and discuss X-ray properties of the LLAGN in
NGC~4579 and origin of the iron emission line.

\section{Observation}
 NGC~4579 was observed on 1995 June 25 with the {\asca} satellite
(\cite{tan94}). The solid state imaging spectrometers (SIS0 and SIS1)
were operated in the 2 CCD Faint and Bright mode. Effective exposure
time was 32~ksec and 3~ksec for Faint and Bright mode data,
respectively, after the standard data screening. As we can correct
echo and dark frame errors (DFE), more reliable determination of
absolute energy is possible for Faint mode data than Bright mode
(\cite{ota94}). Since the SIS sensors were operated in Faint mode
during most of the observation time we used only Faint mode data for
the following analyses. The gas imaging spectrometers (GIS2 and GIS3)
were operated in the pulse-height (PH) nominal mode. We obtained a
useful exposure of 31~ksec after the standard data screening and
rejection of high background time intervals.  X-ray light curves and
spectra were extracted from circular regions centered on NGC~4579 with
a radius of 4' for SIS and 6' for GIS.  The spectra from SIS0 and SIS1
were added together after gain corrections. Spectra from GIS2 and GIS3
were also combined.  Background data were accumulated from a source
free region in the same field. The mean counting rates of SIS and GIS
were 0.15 counts~s$^{-1}$ in 0.5--10~keV and 0.09 counts~s$^{-1}$ in
0.7--10~keV per detector, respectively, after background
subtraction. No significant variability was detected during the
observation. A constant model fit to the light curve with a bin size
of 5760 sec (one orbit of the {\asca} satellite) in the 2-10 keV band
yields reduced chi-square of 1.29 for 16 degrees of freedom. The upper
limit on the variability amplitude is 7\% (r.m.s.) or 26\%
(peak-to-peak) for this light curve.
Therefore we summed all useful data together.

\section{Results}

\subsection{The X-ray Image}

  NGC 4579 is detected at the position of the optical nucleus within
position determination uncertainties and the X-ray image looks
point-like.  We compared X-ray images in the 0.5--2~keV and 2--10~keV
band with the point spread function (PSF) of {\asca} XRT + SIS to
evaluate the spatial extent. We fit the azimuthally averaged
surface brightness profiles with those of a model PSF + constant
background, where we left two parameters free; the normalization of
PSF and the background level.  We
obtained good fits with reduced chi-square $\chi^2_\nu$ = 0.64 and
0.70 (11 dof) for the soft and hard energy bands, respectively.  Thus
the images in these energy bands are consistent with a point
source. In order to set an upper limit to the spatial extent we fit
the radial brightness profiles with those of a two-dimensional
Gaussian convolved through the PSF (see Ptak 1997 for the technique).
In this fitting, the free
parameters are the Gaussian $\sigma$, the normalization of the Gaussian, and
the background level.  The upper limits of the Gaussian width are 0.25
arcmin for both soft and hard band (0.25 arcmin corresponds to 1.2~kpc
at 16.8~Mpc).

The fitted background level is $\sim2.5$ times higher than that of the
blank sky observations released by NASA Guest Observer Facility in the
0.5--2~keV band, while the 2--10~keV background is consistent with the
blank sky fields.  NGC~4579 is located at $\sim1.8^{\circ}$ away from
M87 in the Virgo cluster of galaxies and soft diffuse emission due to
the intracluster gas is present in this region (\cite{boh94}). Thus
the high background level in the soft band is most likely due to Virgo
cluster emission. In the soft band image, no significant structure is
seen.

\subsection{The X-ray Spectrum}

The X-ray spectra obtained with the SIS and GIS are shown in
Fig. 1. The X-ray spectra could not be fitted with simple power-law or
thermal bremsstrahlung model and residuals were clearly seen around
1~keV and 6.7 keV, which can be identified with iron L line complex
and an iron K emission line, respectively. An acceptable fit is
obtained with the sum of a power-law, a Raymond-Smith (R-S) thermal
plasma model (\cite{ray77}) and a Gaussian at 6.7~keV. In the fitting,
the absorption column density is assumed to be the Galactic value
($3.1\times10^{20}$~{\pcm}; \cite{mur96}) for the R-S component and
left free for the power-law component.  The best fit parameters of the
X-ray continuum are summarized in Table~1 and the best fit model are
also shown in Fig. 1 as a histogram. The photon index of the power-law
component is $1.72\pm0.05$ (hereafter quoted errors are 90\%
confidence for one interesting parameter), and the X-ray luminosity of
this component is $1.5\times10^{41}$~{\eps} in the 2--10~keV
band. Although a small excess absorption {\NH} = $(4\pm3) \times
10^{20}$~{\pcm} is necessary to fit the data in addition to the
Galactic absorption, {\NH} is still consistent with the Galactic value
if the calibration uncertainties of the SIS at low energies are taken
into account. Since the power-law component dominates the X-ray flux
even in the soft energy band, the abundance of the metals in the R-S
component is poorly constrained. Therefore we fixed the abundance at
0.5 solar, which is typical for hot gas in spiral galaxies (e.g. Ptak
1997). Then the soft component is represented by the R-S model with
$kT=0.90^{+0.11}_{-0.05}$ keV. The addition of the R-S component to
the power-law + Gaussian model improved chi-square significantly
($\Delta\chi^2$=33.4) and this decomposition of the spectrum is very
similar to that of many other low luminosity objects observed by
{\asca} (\cite{ser96}). The X-ray luminosity of the R-S component is
$1.2 \times10^{40}$~{\eps} in the 0.5--4~keV band. The luminosity of
the R-S component depends on the assumed abundance value, and
$0.88\times10^{40}$ and $1.7 \times10^{40}$~{\eps} are obtained for
assumed abundance values 1.0 and 0.1 solar, respectively. The hard
band spectrum can be also represented by a Raymond-Smith thermal
plasma with $kT=7.9^{+1.3}_{-0.9}$~keV and abundance of
$0.55^{+0.18}_{-0.16}$ instead of a power-law plus Gaussian
(Table~1).

\placefigure{fig1}

\placetable{tbl-1}

  An iron K line is clearly detected in the X-ray spectra: the
addition of a Gaussian line improved $\chi^2$ by 20.0 for three
additional parameters (line center energy, line width, and
normalization).  Therefore the iron line is statistically significant
at more than 99.9\% confidence according to the F-test. The line
center energy is $6.73^{+0.13}_{-0.12}$~keV (rest frame), which is
higher than the 6.4~keV typically observed from Seyfert 1 galaxies.
The equivalent width is $490^{+180}_{-190}$~eV. The iron line profile
is shown in Figure 2 as the ratio of the data to the best fit
continuum component of the above fits. Figure 3 shows the
confidence contours for the line energy and intensity. The best fit
energy agrees with He-like iron and the 90\% confidence range
corresponds to the ionization state from \ion{Fe}{20} to \ion{Fe}{25} 
(He-like). A
line center energy of 6.4~keV, which is expected from cold or low-ionization
iron, is excluded at more than the 99\% confidence level. The confidence
contours for the line width $\sigma$ and intensity are shown in Figure 4. The emission line is marginally broad. Although the
best fit width is $\sigma$ = 170~eV, a narrow line cannot be excluded
at 90\% confidence level for two interesting parameters. If we fix the
line width at $\sigma$=0, a line center energy of
$6.82^{+0.10}_{-0.26}$~keV and equivalent width of
$360^{+175}_{-135}$~eV are obtained.

A combination of multiple narrow lines instead of a single broad
Gaussian also provides a good fit ($\chi^2_{\nu}$=0.958 for 201 dof),
where the line center energies are fixed at 6.4~keV, 6.7~keV, and
7.0~keV, which represent cold, He-like, and H-like iron, respectively.
The obtained equivalent widths are $110^{+130}_{-110}$,
$240^{+170}_{-160}$, and $160^{+190}_{-160}$~eV, respectively (Table
2). The 6.7~keV line is the dominant component also in this model.

Many Seyfert 1 galaxies show broad iron lines with a significant red
tail and they are interpreted as originating from the inner part of a
relativistic accretion disk (e.g. Tanaka et al. 1995). We examined the
disk-line model by Fabian et al. (1989) instead of the Gaussian
model. Since the statistics are limited, only two parameters, the
inclination angle of the disk and normalization, were left free. The
inclination angle is defined such that $i=0$ corresponds to a face-on
disk. The line emissivity is assumed to be proportional to $r^{-q}$
and $q$ is fixed at 2.5, which is the typical value for Seyfert 1 galaxies
(Nandra et al. 1997a). The line center energy is fixed at 6.4~keV or
6.7~keV. The inner radius of the line-emitting region was fixed at
6$r_g$, where $r_g = GM/c^2$ is the gravitational radius.  The outer
radius is fixed at the $\chi^2$ local minima for the fits, 16.6$r_g$ and
10.5$r_g$ for 6.4~keV and 6.7~keV case, respectively. The disk-line
model fits provided worse reduced $\chi^2$ values
$\Delta\chi^2\sim$8 than the Gaussian modeling of the line. The best fit
parameters are summarized in Table 3. Although the fit is acceptable,
systematic positive residuals are seen around 6.7~keV. This is
probably due to absence of significant red asymmetry in the observed
profile. The obtained equivalent width $\sim 900$~eV is extremely
large compared to the value expected from a X-ray irradiated disk
(e.g. \cite{geo91}) and observed in Seyfert 1 galaxies
(Nandra et al. 1997a).

\section{Discussion}

\subsection{X-ray emission from a low luminosity AGN}

  We obtained X-ray images and spectra in the 0.5--10~keV band and a
point-like X-ray source with a photon index of $\Gamma = 1.72\pm0.05$
is detected. An iron line is also detected at 6.7~keV. In the soft
energy band, a broad line like feature identified with iron-L line
complex indicates the presence of thin-thermal plasmas of temperature
$kT\sim 0.9$~keV.

  The X-ray luminosity ($1.5\times 10^{41}~${\eps} in 2--10~keV ) is
1--3 orders of magnitude smaller than typical Seyfert galaxies and
falls in the classes of LINERs and "low luminosity" Seyfert galaxies
(Serlemitsos et al. 1996, Iyomoto et al. 1996, Ishisaki et al. 1996,
Terashima et al. 1998b). In normal spiral galaxies, the X-ray emission
is dominated by discrete sources, specifically low mass X-ray binaries
(LMXBs) (e.g. \cite{fa89}, Makishima et al. 1989). The X-ray
luminosity from LMXBs are roughly proportional to B-band luminosity
{\LB} and their X-ray spectra can be approximated by a thermal
bremsstrahlung of a temperature of several keV. The {\asca} X-ray
spectrum of NGC~4579 is also fitted by $kT\sim8$~keV thermal plasma
model. However the strong iron line at 6.7~keV is not compatible with
the X-ray spectra of LMXBs, since the equivalent width of iron
emission lines from LMXBs are small (several tens of eV,
\cite{hir87}). Additionally, the {\LX}/{\LB} value $1.3\times10^{-3}$
is more than an order of magnitude higher than normal spiral galaxies
(e.g. \LX /\LB =$3.5\times10^{-5}$ for M31;
\cite{mak89}). Additionally an upper limit on the size of an archival
{\rosat} PSPC image is 14" (Gaussian $\sigma$), which corresponds to
1.1 kpc at 16.8 Mpc. This upper limit is significantly lower than the
size of the galaxy.  Therefore we conclude that contribution from
LMXBs to X-ray emission of NGC~4579 is negligible.

  Hot plasmas with temperatures on the order of $\sim10$ keV are present in the
Galactic center region and their X-ray spectra show prominent, ionized
iron K emission (e.g. Koyama et al. 1996).  The X-ray spectral shape
of NGC 4579 in the hard X-ray band is similar to such hot
gas. However, the X-ray luminosity of NGC 4579 is three orders of
magnitude higher than the Galactic ridge emission
({\LX}$\sim2\times10^{38}$ {\eps}; Kaneda et al. 1997, Yamasaki 1996,
Warwick et al. 1985).  Starburst galaxies also show a hard spectral
component with a temperature of $\sim$10 keV and their X-ray
luminosities are around $10^{40}$ {\eps} (e.g. $3.4\times10^{40}$ \eps
in 2--10 keV for M82; Ptak et al. 1997). However the starburst
activity in NGC 4579 is weaker than M82, since the far-infrared
luminosity of NGC 4579 is about an order of magnitude lower than that
of M82. Furthermore, starburst galaxies show weak or no iron emission
contrary to NGC 4579. Therefore hot plasma is unlikely as the origin
of the hard component and iron emission line in NGC 4579 and we
conclude that the AGN emission dominates the {\it ASCA} spectra and that
other components such as a hot gas contribution is small, if any.

We note that errors in background subtraction of the Virgo cluster hot
gas do not affect the detection of the iron emission line at 6.7 keV,
since the cluster gas is very dim in this region and temperature is
low ($kT\sim2$ keV; Matsumoto 1998, \cite{boh94}). Actually no
significant iron emission is detected from the GIS field around NGC
4579.

  If the primary ionizing mechanism of LINER optical emission lines in
this galaxy is photoionization by a LLAGN, {\LX}/{\LHa}
might be expected to be similar to Seyfert 1 galaxies, for which there is a
good positive correlation between {\LX} and {\LHa} (e.g. Ward et
al. 1988, \cite{kor95}, Serlemitsos et al. 1996). Using the
H$\alpha$ luminosity of broad plus narrow component {\LHa} =
$5.9\times10^{39}$~{\eps} (\cite{ho97b}) and the observed X-ray
luminosity in the 2--10~keV band, we obtain {\LX}/{\LHa} $\approx$ 26
for NGC~4579. This value is in excellent agreement with those of
Seyfert 1 galaxies (\cite{war88}) and strongly supports a low
luminosity AGN as the ionizing source of the LINER in NGC~4579.

  Less luminous Seyfert 1 galaxies tend to show rapid and large
amplitude variability (Mushotzky, Done, \& Pounds 1993 and references
therein). However NGC 4579 shows no significant short term
variability. Lack of variability on short time scales seems to be a
common property of LLAGNs (\cite{mus92}, \cite{pet93}), for example
the LLAGN in NGC 1097 (Iyomoto et al. 1996) and NGC 3998
(\cite{awa92}) also show no significant variability on timescales less
than a day. Direct comparison of {\rosat} PSPC and {\asca} flux in the
0.5--2 keV band show a factor of two increase in $\sim$3.5 years.

  The X-ray spectral slope $\Gamma = 1.72\pm0.05$ is identical to the
average value found for hard X-ray selected Seyfert 1 galaxies
(Mushotzky et al. 1993) but the luminosity is lower than that of any
Seyfert 1 galaxy but NGC 4051. Based on the FW0I (full width at 0
intensity) of a broad emission line and an estimate of the size of the
broad line region, mass of the central black hole is roughly estimated
to be
 $M_{\bullet} \sim 4\times10^6M_{\odot}$. Then the Eddington ratio
$L/L_{\rm Edd}$ is $\sim 10^{-3}$ for the observed luminosity of
$\sim5\times10^{41}$~{\eps} (Barth et al.  1996), although their
blackhole mass estimation is crude. Therefore the X-ray spectral slope
does not seem to be drastically changed even at a very low Eddington
ratio. This is also true for M81, for which $L/L_{\rm Edd}$ is
estimated to be $\sim (2-10)\times 10^{-4}$ (\cite{ho96}) and the
photon index is $1.85\pm0.04$ (\cite{ish96}).

Soft thermal emission of $kT\sim0.5-1$~keV is often observed from low
luminosity AGNs (Terashima 1997, \cite{pta97}, Serlemitsos et
al. 1996). In some cases, such emission is associated with starburst
activity (e.g. \cite{iyo96}, \cite{te98a}).  Since the far-infrared
luminosity of NGC~4579 is $1.5\times10^{43}$~{\eps} some star
formation activity may be present which may explain the thermal
emission. The soft thermal X-ray to far infrared luminosity ratio
{\LX}/{\LFIR} = $6\times10^{-4} - 1.1\times10^{-3}$ is consistent with
starburst galaxies (e.g. \cite{dav92}) within the scatter.

\subsection{Iron-K line}

  A marginally broad ($\sigma \approx 0.17$~keV) iron emission line is
clearly detected at $6.73^{+0.13}_{-0.12}$~keV and the equivalent
width is $490^{+180}_{-190}$~eV for the broad Gaussian model fit. The
line center energy is significantly higher than 6.4~keV, which is
typically observed from Seyfert 1 galaxies, and consistent with
He-like iron.  A similar broad iron line centered at $\sim 6.7$~keV is
detected from the low luminosity Seyfert galaxy M81 (Ishisaki et
al. 1996, Serlemitsos et al. 1996). The line can also be represented
by line blending of neutral, He-like, and H-like iron and dominated by
He-like iron.  The disk-line profile (Fabian et al. 1989) is probably
inconsistent with the data for 6.4~keV or 6.7~keV intrinsic line
energy because of following reasons. The $\chi^2$ value is worse
than a single broad Gaussian fit and systematic residuals remain in
the disk-line fit, since a significant red tail is not clearly seen in
the data.  Furthermore, the disk-line model provides the very large
equivalent width $\sim900$~eV, which is about 4 times larger than the
results of the disk-line fit to Seyfert 1 galaxies
($<$EW$>=(230\pm60)$ ~eV, \cite{na97a}). Therefore our data prefer a
symmetric Gaussian-shape profile with intrinsic line center energy of
6.7~keV (He-like) rather than 6.4~keV ($<$\ion{Fe}{16}). Thus the
ionization state of the iron line emitter may be different from that
of higher luminosity Seyfert 1 galaxies in at least some LLAGNs (NGC
4579 and M81)

Strong ionized iron emission lines are observed in heavily obscured
Seyfert 2 galaxies (NGC 1068, \cite{uen94}, \cite{iwa97}; NGC 1365,
\cite{iyo97}; see also \cite{tu97a}, \cite{tu97b}). In these objects
continuum emission from the nucleus is completely blocked and only
scattered radiation is observed.  Ionized iron lines are interpreted
as originating from a photoionized scattering medium. If the
continuum of NGC~4579 is scattered radiation, then the observed X-ray
luminosity is only a fraction of its intrinsic luminosity. Since the
scattering fraction is typically less than 10 \% for Seyfert 2
galaxies (\cite{uen95}), the {\LX}/{\LOIII} should be less than 10 \%
of those of Seyfert 1 galaxies as is the case for NGC 1068
(\cite{mul94}).  However the observed X-ray to [OIII]$\lambda$5007
luminosity ratio {\LX}/{\LOIII} is very similar to Seyfert 1 galaxies. 
Therefore the observed X-ray continuum is not likely to be due to
a scattered component. Then the observed iron line should be emitted
from the matter close to the nucleus in order to be ionized and/or
broadened due to the Doppler effect.

If the iron line is emitted by an accretion disk, a line profile with
significant red tail is expected (\cite{fab89}). On the other hand,
the observed profile seems to be symmetric in shape although the
statistics are limited. Broad lines with weaker red tails than Seyfert
1s are observed in AGNs with much higher luminosity;
{\LX}$>10^{44}$~{\eps} (\cite{na97b}). If the inner-most part of the
accretion disk is almost fully ionized, the red component is expected
to be very weak or absent. Thus the observed profile is consistent with
the interpretation that the observed iron K emission is from an
ionized disk.

The obtained equivalent width ($\sim 500$~eV for the Gaussian model)
is rather large compared to that seen in most Seyfert 1 galaxies.  If the
disk is highly ionized, the fluorescence yield of iron increases and
absorption
by lighter elements decreases as light elements are almost completely
ionized.  In such a situation the equivalent width of an iron
line can increase by a factor of two (Matt et al. 1993, \cite{zyc94}).
Therefore the large equivalent width is also naturally explained by an
ionization effect.

The ionization state of photoionized matter is determined by an
ionization parameter $\xi = L/nR^2$ (\cite{kal82}), where
$L$, $n$, and $R$ is the luminosity of ionizing photons, the number density of
photoionized matter, and the distance from light source to photoionized
matter, respectively. The X-ray luminosity of NGC 4579 is only
$1.5\times10^{41}$ {\eps}, which is 1--3 orders of magnitude smaller
than for Seyfert 1 galaxies, and the X-ray luminosity of M81, from
which an iron line centered at $\sim6.7$~keV is detected, is even
lower ($\sim 2\times10^{40}$ {\eps}). In order to photoionize iron
atoms to be He-like, $\xi$ should be at least $\sim$ 500, while $\xi<100$
is required for less ionized species ($<$ \ion{Fe}{16}) which is probably
appropriate for usual Seyfert 1 galaxies. Therefore $nR^2$ in the iron
line emitting region should be more than two orders of magnitude
smaller than that of luminous Seyfert 1 galaxies. An expected
ionization parameter under an assumption of standard $\alpha$ disk is
calculated by \cite{mat93}. According to their results, the
ionization parameter has a strong dependence on the mass accretion
rate $\xi \propto \dot{m}^3$ (equations (5) and (6) in Matt et
al. 1993), where $\dot{m}$ is denoted in units of the critical
accretion rate $\dot{m} = L/L_{\rm Edd}$. In order to ionize iron to
He-like, $\dot{m}$ should be at least 0.2 (Figs. 2 and 5 in Matt et
al. 1993). However the order-of-magnitude estimate of the central
black hole mass by \cite{bar96} combined with the observed
luminosity gives a significantly smaller value of $\dot{m}$
$\sim1\times10^{-3}$. Then we cannot explain the very low luminosity
and the ionized iron line at the same time in the standard disk model. 
This may suggest that the accretion processes in AGN is different in a
very low luminosity situations with very small $\dot{m}$.

  An advection dominated accretion flow (ADAF) model is proposed for
AGNs specifically for objects radiating at very low Eddington ratio
(e.g. $\dot{m} \sim 10^{-4}$ for NGC 4258, Lasota et al. 1996). In the
model by Lasota et al. (1996), a standard disk is assumed outside of
$r_{\rm in}$ and an ADAF inside of $r_{\rm in}$. In an ADAF, accreting
matter is heated up to very high temperatures ($T_i\sim10^{12}$K,
$T_e\sim10^{9}$K).  However our detection of an iron line indicates
the presence of highly ionized (but not fully ionized)  matter
surrounding a large solid angle viewed from the light source.  This
means that $r_{\rm in}$ should be small and a geometrically thin disk is
appropriate. Therefore the iron line in NGC~4579 cannot be explained
solely by an ADAF model and the real situation in NGC~4579 may
correspond to a condition near the transition from the $\alpha$ disk to
an ADAF.

Future sophisticated modeling of accretion in LLAGNs and calculation
of expected iron emission as well as precise measurements of an iron K
line and mass determination by {\HST} Space Telescope Imaging
Spectrograph will be important to understand physical processes in
extremely low luminosity AGNs.\\

\section{Summary}

We observed the LINER NGC~4579 with {\asca} and detected X-ray
emission with a luminosity $1.5\times10^{41}$~{\eps} probably from a LLAGN.
The X-ray spectral slope ($\Gamma=1.72\pm0.05$) is quite similar to
Seyfert 1 galaxies. Iron K
emission is detected at 6.7~keV, which is consistent with He-like
iron, with an equivalent width of 500~eV. Although the statistics are limited,
the observed iron line profile shows no significant red tail and
a  symmetrically-shaped He-like iron line is preferred rather than a
blueshifted component of the disk-line profile from neutral or low-ionization
iron as seen in Seyfert 1 galaxies .
The observed center energy, profile, and equivalent width of iron
emission are well explained in terms of an ionized disk
origin. However the ionization parameter cannot be large enough to ionize
the disk with the inferred low Eddington ratio (Matt et
al. 1993). The detection of iron emission indicates that a large solid
angle seen from the X-ray source is surrounded by
ionized material. This is inconsistent with an advection dominated
accretion flow model for LLAGNs (\cite{las96}).\\

\acknowledgments

The authors are grateful to all the {\asca} team members. YT and KM thank
JSPS for support.

\clearpage
 
\begin{table*}
\begin{center}
\begin{tabular}{cccccccc}
\tableline \tableline
model   & \NH (galactic) & $kT$                 & abundance     & \NH
& $\Gamma$ or $kT$~[keV]     &  abundance       & $\chi^2$/dof\\
        & [$10^{20}$~{\pcm}]    & [keV]         & [solar]       &
[$10^{20}$~{\pcm}]    &       & [solar]   &\\ 
\tableline
(1)     & 3.1(f)      & $0.90^{+0.11}_{-0.05}$ & 0.50(f)        & $4.1\pm2.7$ &
$1.72\pm0.05$         &  ---    & 192.4/201 \\
(2)     & 3.1(f)      & $0.88^{+0.11}_{-0.05}$ & 0.50(f)        & 0(f) 
& $7.9^{+1.3}_{-0.9}$ & $0.55^{+0.18}_{-0.16}$  & 198.2/204 \\
\tableline
\end{tabular}
\end{center}
\tablenum{1}
\caption{Results of spectral fitting to the SIS and GIS spectra of NGC4579\label{tbl-1}}
\tablecomments{The fitting models are (1)Raymond-Smith + Power-law and
(2)Raymond-Smith + Raymond-Smith. (f) in the table denotes frozen
parameter. The quoted errors in parenthesis are at the 90\% confidence
level for one interesting parameter. }
\end{table*}


\begin{table*}
\begin{center}
\begin{tabular}{ccccc}
\tableline \tableline
model           & $E_L$         & $\sigma$      & EW 	& $\chi^2$/dof\\
                & [keV]         & [keV]         & [eV]	& \\
\tableline
narrow Gaussian	& $6.82^{+0.10}_{-0.26}$ & 0(f) & $360^{+175}_{-135}$	&
195.5/202\\
broad Gaussian  & $6.73^{+0.13}_{-0.12}$ & $0.17^{+0.11}_{-0.12}$ &
$490^{+180}_{-190}$ & 192.4/201\\
three narrow Gaussians  & 6.4(f) & 0(f)       & $110^{+130}_{-110}$ & 192.6/201\\
			& 6.7(f) & 0(f)       & $240^{+170}_{-160}$ &\\
                        & 7.0(f) & 0(f)       & $160^{+190}_{-160}$ &\\
\tableline
\end{tabular}
\end{center}
\tablenum{2}
\caption{Gaussian fits to the iron K line. \label{tbl-2}}
\end{table*}


\begin{table*}
\begin{center}
\begin{tabular}{ccccccc}
\tableline \tableline
        & $E_L$         & $R_{in}$ & $R_{out}$  & $i$   & EW   &$\chi^2$/dof\\ 
        & [keV]         & [$R_g$]  & [$R_g$]    & degree & [eV] & \\
\tableline
        & 6.4(f)         & 6(f) & 16.6(f)       & $42\pm4$      &
$880^{+570}_{-360}$ & 200.7/202\\
        & 6.7(f)         & 6(f) & 10.5(f)       & $36\pm3$      &
$920^{+530}_{-320}$ & 201.5/202\\
\tableline
\end{tabular}
\end{center}
\tablenum{3}
\caption{Disk-line model fits to the iron K line \label{tbl-3}}
\end{table*}

\clearpage


\figcaption{{\asca} SIS and GIS spectra of NGC 4579. 
The best fit model consist of a Raymond-Smith thermal plasma, power-law, 
and Gaussian is shown as histograms. SIS(a) and GIS(b) data are plotted
separately for clarity.    \label{fig1}}


\figcaption{Data/model ratio for the best fit continuum 
model around an iron K emission line. 
The crosses with and without filled circle represent SIS and GIS data, 
respectively. Energy scale is not redshift corrected.
\label{fig2}}



\figcaption{Confidence contours for the line energy 
and intensity. Energy scale is redshift corrected.
The contours correspond to 68\%, 90\%, and 99\% 
confidence level for two interesting parameters 
($\Delta \chi^2$ = 2.3, 4.6, and 9.2, respectively). 
\label{fig3}}


\figcaption{Confidence contours for the line width 
(Gaussian $\sigma$) and intensity. 
The contours correspond to 68\%, 90\%, and 99\% confidence
level for two interesting parameters. \label{fig4}}


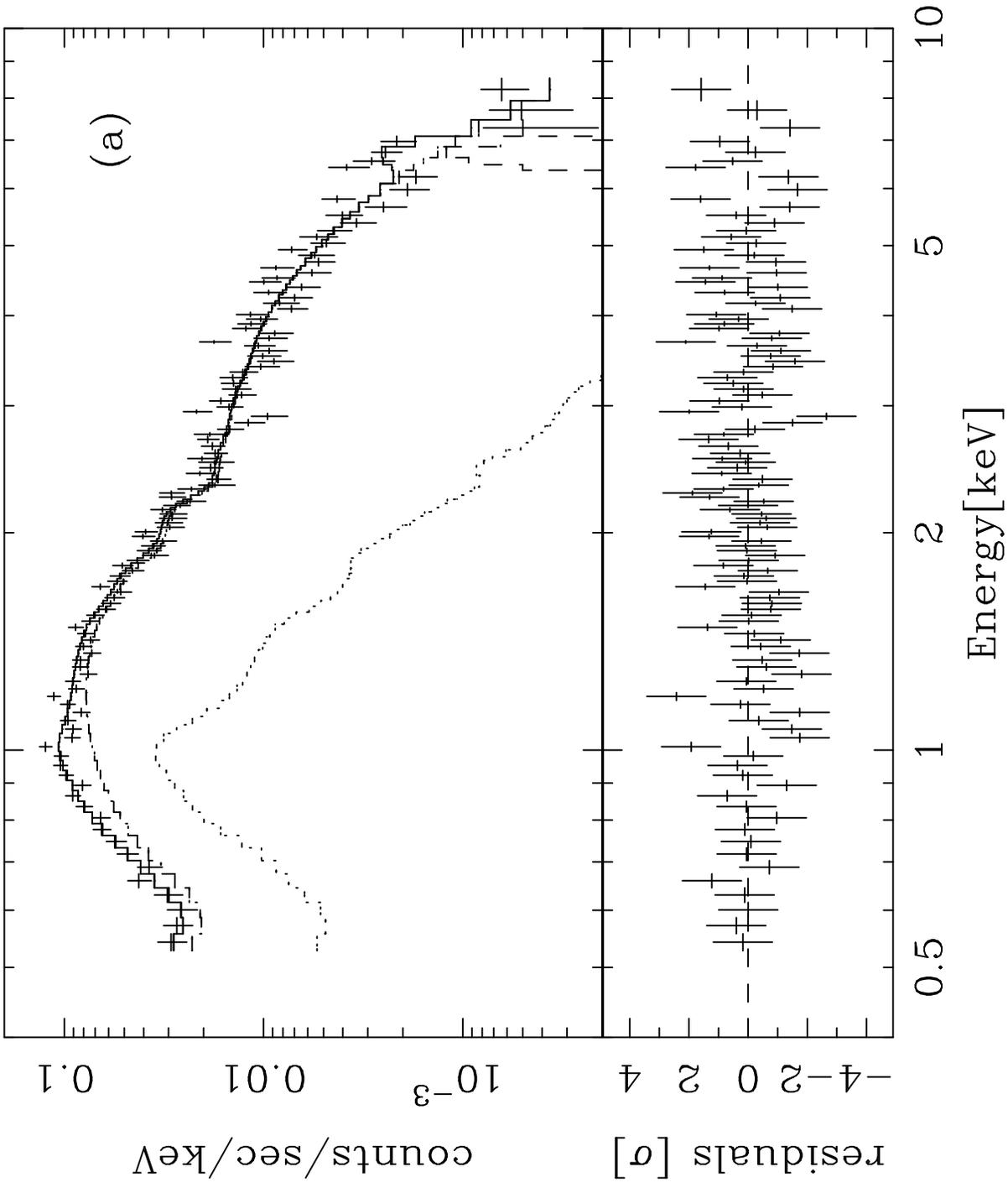
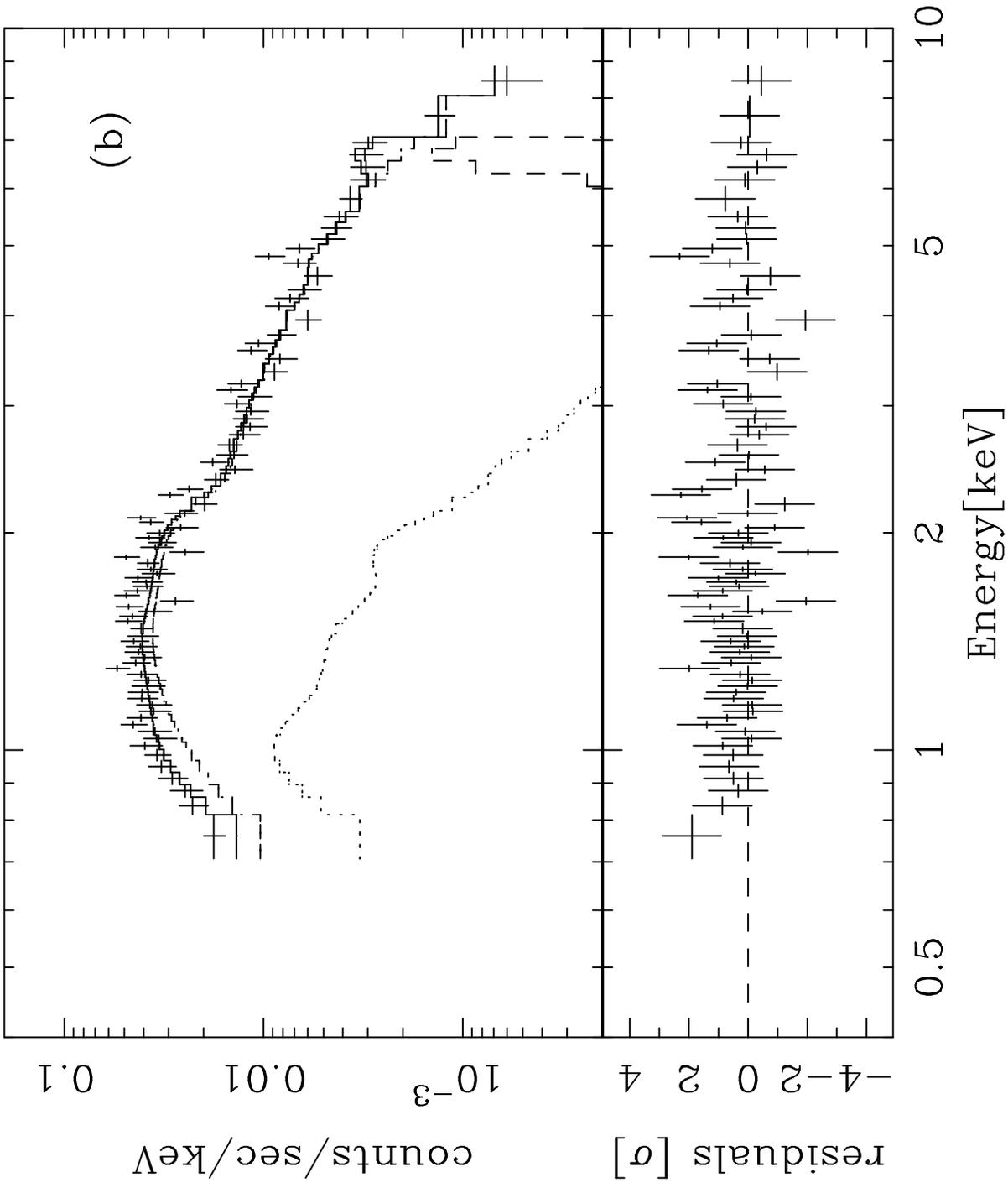
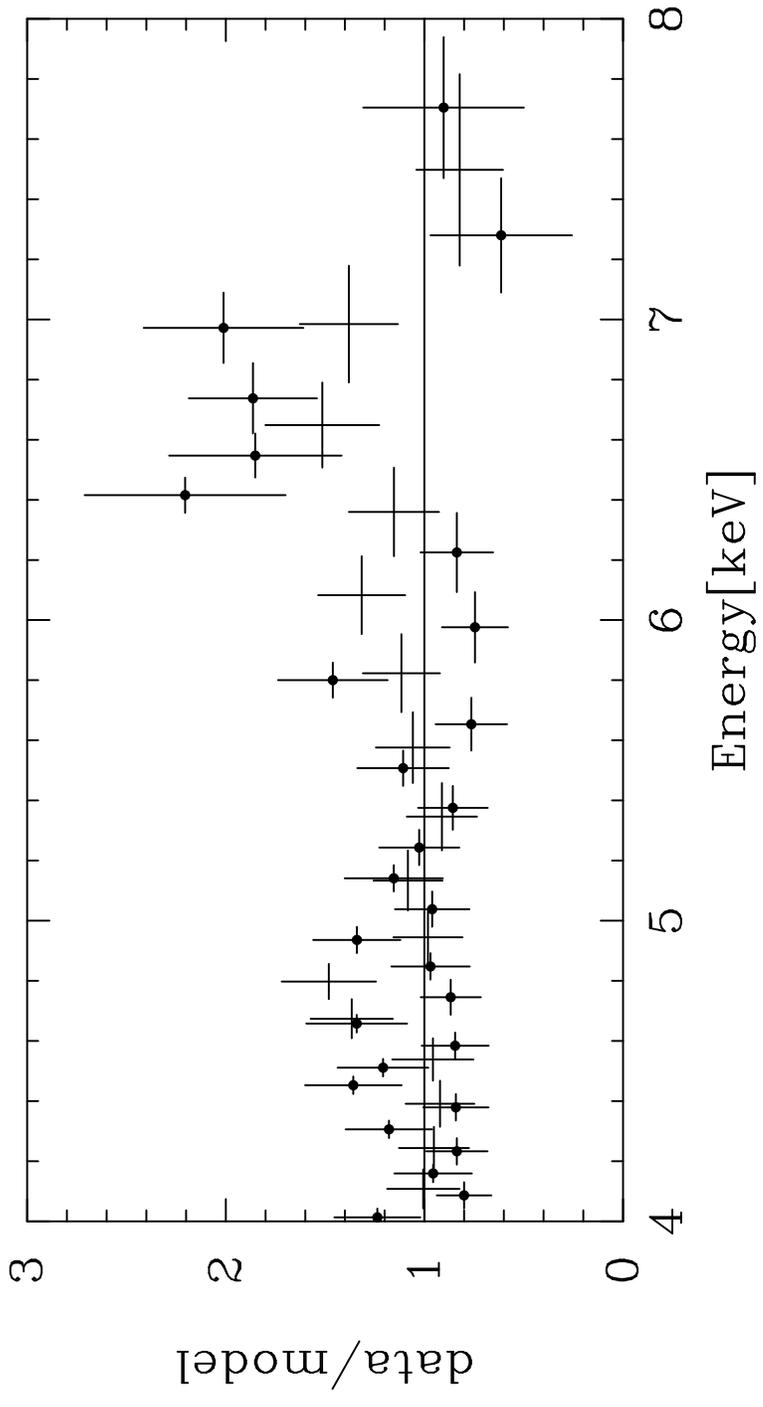
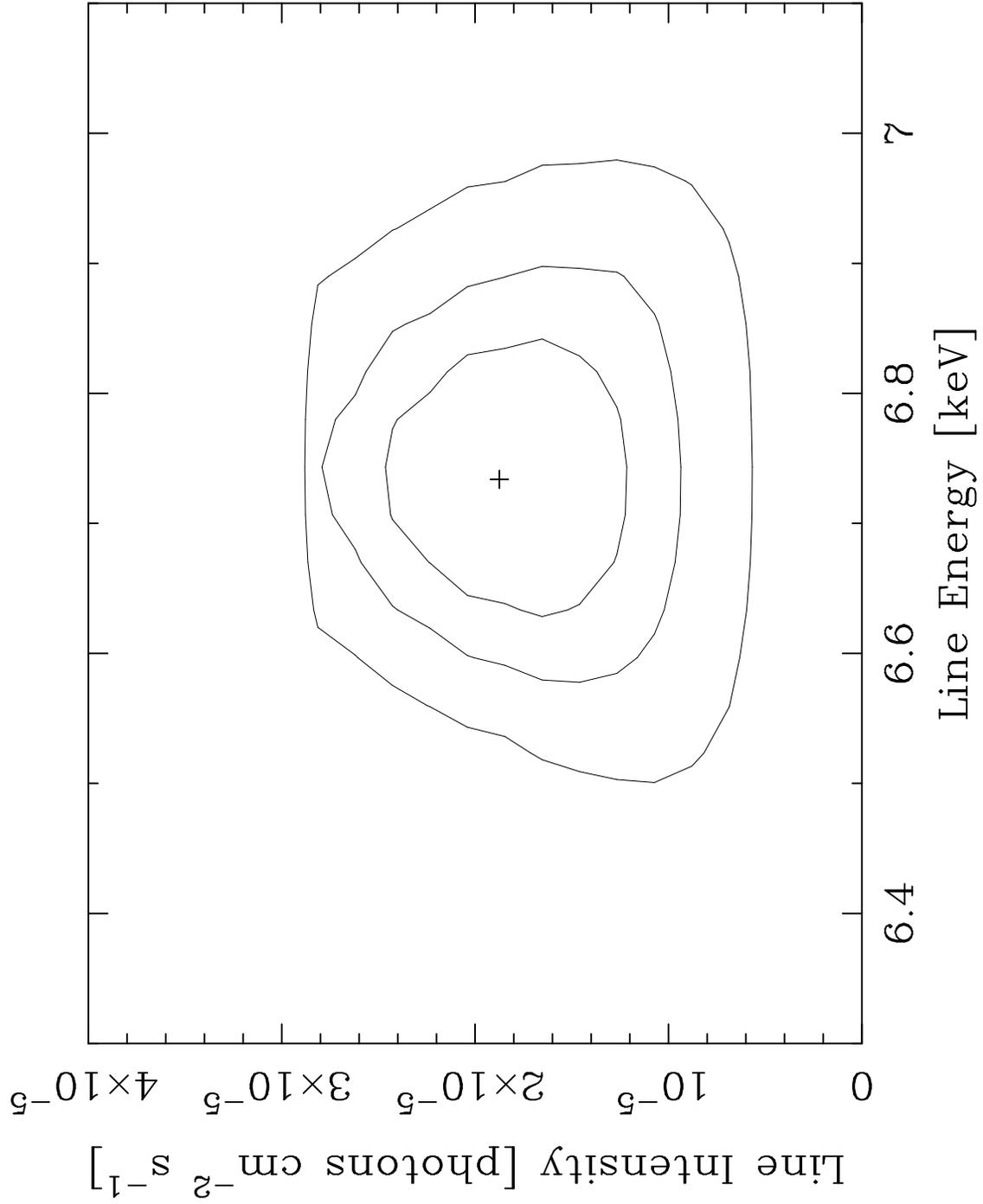
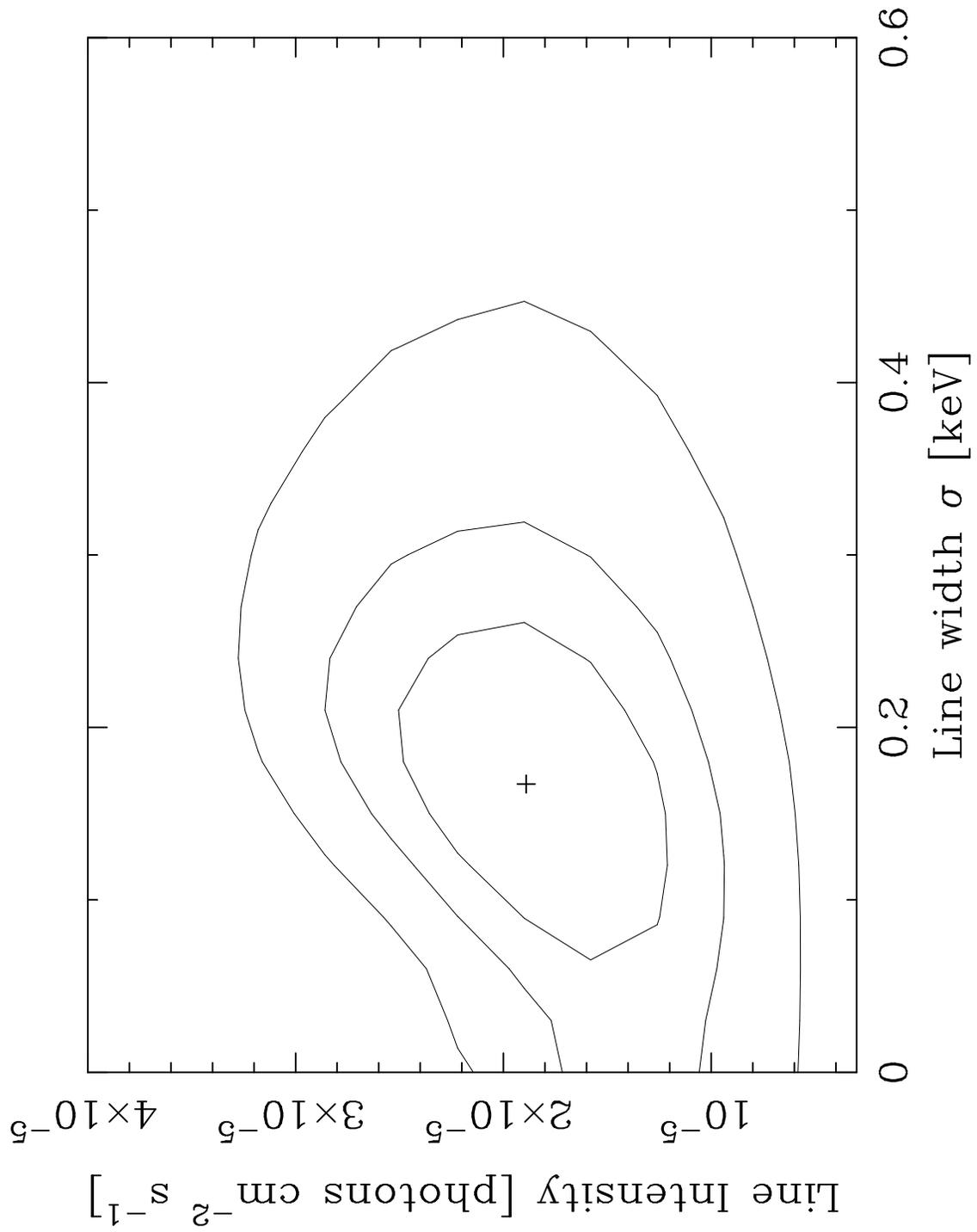
\end{document}